\documentstyle[aps]{revtex}

\draft

\begin{document}

\tighten
\twocolumn[\hsize\textwidth\columnwidth\hsize\csname@twocolumnfalse\endcsname

\title{Two-channel Kondo model as a generalized one-dimensional inverse
square long-range Haldane-Shastry spin model}
\author{Guang-Ming Zhang$^1$and Lu Yu$^{2,3}$}

\address{$^1$Center for Advanced Study, Tsinghua University, Beijing 100084,
P. R. China.\\
$^2$International Center for Theoretical Physics, P. O. Box 586, Treiste
34100, Italy\\
$^3$Institute of Theoretical Physics, CAS, Beijing 100080, P. R. China}

\date{\today}

\maketitle

\begin{abstract}
Majorana fermion representations of the algebra associated with spin,
charge, and flavor currents  have been used to transform the two-channel
Kondo Hamiltonian. Using a path integral formulation, we derive
a reduced effective action with long-range impurity spin-spin interactions
at different imaginary times. In the semiclassical limit, it is
equivalent to a one-dimensional Heisenberg spin chain with two-spin,
three-spin, etc. long-range interactions, as a generalization of the
inverse-square long-range Haldane-Shastry model. In this representation
the elementary excitations are "semions", and the non-Fermi-liquid
low-energy properties of the two-channel Kondo model are recovered.

\end{abstract}

]

The two-channel Kondo model is known to have a non-Fermi liquid (non-FL)
low-energy fixed point in the overscreening case, and has been put forward
as a model to explain non-FL behavior observed in several quite different
physical systems at low temperatures, such as certain heavy fermion alloys
and two-level systems \cite{cz}. There are exact solutions for the ground
state and thermodynamics of this model derived from the Bethe Ansatz \cite
{andrei,schlottmann}, but there are still continuing efforts to find an
intuitive understanding of the nature of excitations in the neighborhood of
the low-energy fixed point. Numerical renormalization group \cite{pc},
conformal field theory (CFT) \cite{al}, and bosonization approach \cite
{ek,ml,cit} made a number of predictions for the many-body excitations at
the fixed point, but they do not provide an intuitive interpretation of
these excitations as it is possible at the FL fixed point of the
single-channel Kondo model \cite{wilson}.

It was shown a long time ago that the single-channel Kondo model can be
reduced to an inverse-square one-dimensional {\it Ising} model \cite{ay},
which is a prototype classical model in statistical physics \cite
{dyson,cardy}. Moreover, such a reduced effective model had helped Anderson
and coworkers to establish the correct FL behavior of the low-energy fixed
point for the one-channel model in the early 70s \cite{ay,ayh}, namely, the 
{\it Ising} spin-spin correlation function should behave as $\frac 1{\tau ^2}
$ with $\tau $ as the imaginary time. On the other hand, the non-FL behavior
of the two-channel Kondo model is characterized by the dynamic correlation
function of the impurity spin, $<T_\tau \overrightarrow{S_d}(\tau )\cdot 
\overrightarrow{S_d}(0)>\sim \frac 1{\mid \tau \mid },$ but its physical
meaning, unfortunately, has not been fully understood yet. To our knowledge,
the possible connection of the overscreened two-channel Kondo model with
quantum spin models has not been explored so far.

In this Letter, the algebra of total spin, charge, and flavor currents of
the two-channel Kondo model have been represented in terms of Majorana
fermions, and in the path integral formulation, we derive a reduced
equivalent quantum Heisenberg spin model, in which the impurity spins at
different imaginary times are strongly correlated, including two-spin,
three-spin, etc. long-range exchange interactions. In particular, the
two-spin interaction has exact the same form as the integrable
inverse-square one-dimensional Heisenberg spin chain --- the so-called
Haldane-Shastry (HS) model \cite{hs}, while the three-spin and four-spin
interaction parts, etc. are natural generalizations of the HS model. The
non-FL fixed point action of the two-channel Kondo model is identified with
the two-spin interaction part (the HS model), and the elementary excitations
of this low-energy non-FL fixed point are spinons ($S=1/2$ objects) obeying
semion (half-fractional) statistics intermediate between bosons and fermions 
\cite{haldane}. Actually, the two-spin long-range interaction part forms 
{\it an ideal semion gas}, while the three-spin long-range term is a {\it %
dangerous irrelevant interaction}, leading to important corrections to the
thermodynamic properties. The fourth and higher order terms are irrelevant
variables. The non-FL properties of the two-channel Kondo model are thus
recovered.

We start with the Hamiltonian of the two-channel Kondo model in the form, 
\begin{eqnarray}
&&H=H_0+H_I  \nonumber \\
&&H_0=\frac{v_f}{2\pi }\sum_{j=1}^2\sum_{\sigma =\uparrow ,\downarrow
}\int_{-\infty }^{+\infty }dx\text{ }\psi _{j,\sigma }^{\dag }(x)(i\partial
_x)\psi _{j,\sigma }(x)  \nonumber \\
&&H_I=\sum_{a=x,y,z}J_aS_d^aJ_s^a(0),
\end{eqnarray}
where we have retained only the s-wave scattering, linearized the fermion
spectrum, and replaced the incoming and outgoing waves with two left-moving
electron fields $\psi _{j,\sigma }(x)$. $J_s^a(x)$ are the conduction
electron spin current operators: $J_s^a(x)=\sum_{j,\sigma ,\sigma ^{\prime
}}:\psi _{j,\sigma }^{\dag }(x)s_{\sigma ,\sigma ^{\prime }}^a\psi
_{j,\sigma ^{\prime }}(x):$, where $s^a$ being spin-1/2 matrices, :: means
normal ordering. We introduce charge and flavor currents: $%
J_c(x)=\sum_{j,\sigma }:\psi _{j,\sigma }^{\dag }(x)\psi _{j,\sigma }(x):$
and $J_f^a(x)=\sum_{j,j^{\prime },\sigma }:\psi _{j,\sigma }^{\dag
}(x)t_{j,j^{\prime }}^a\psi _{j^{\prime },\sigma }(x):$, where $%
t_{j,j^{\prime }}^a$ are generators of an SU(2) symmetry group. Following
Affleck and Ludwig \cite{al}, the free part of the Hamiltonian can be
rewritten as a sum of three commuting terms by the usual point-splitting
procedure (Sugawara construction): 
\begin{eqnarray}
H_0=\frac{v_f}{2\pi }\int_{-\infty }^{+\infty }dx &&\left[ \frac 18%
:J_c(x)J_c(x):+\frac 14:\vec{J}_f(x)\cdot \vec{J}_f(x):\right.  \nonumber
\label{suga} \\
&&\left. +\frac 14:\vec{J}_s(x)\cdot \vec{J}_s(x):\right] ,
\end{eqnarray}
while the interaction term is expressed only in terms of the electron spin
currents and the impurity spin. The information about the number of channels
is contained in the commutation relations obeyed by the spin currents 
\[
\lbrack J_s^a(x),J_s^b(x^{\prime })]=i\epsilon ^{abc}J_s^a(x)\delta
(x-x^{\prime })+\frac{ki}{4\pi }\delta _{a,b}\delta ^{\prime }(x-x^{\prime
}) 
\]
indicating that $J_s^a(x)$ form an SU(2) level $k=2$ Kac-Moody algebra.
Meanwhile, the charge and flavor currents satisfy 
\begin{eqnarray*}
\lbrack J_c(x),J_c(x^{\prime })] &=&2ki\delta ^{\prime }(x-x^{\prime }), \\
\lbrack J_f^a(x),J_f^b(x^{\prime })] &=&i\epsilon ^{abc}J_f^a(x)\delta
(x-x^{\prime })+\frac{ki}{4\pi }\delta _{a,b}\delta ^{\prime }(x-x^{\prime
}).
\end{eqnarray*}
They form a U(1) Kac-Moody and another SU($k=2$) level-2 Kac-Moody algebra,
respectively.

It is now quite natural to introduce a Majorana representation of the spin
current operators, 
\begin{eqnarray}
&&J_s^x(x)=-i\chi _2(x)\chi _3(x),  \nonumber \\
&&J_s^y(x)=-i\chi _3(x)\chi _1(x),  \nonumber \\
&&J_s^z(x)=-i\chi _1(x)\chi _2(x),
\end{eqnarray}
where $\chi _1(x),\chi _2(x),$ and $\chi _3(x)$ are left-moving free
Majorana fermion fields, and it can be shown to reproduce the SU(2) level-2
Kac-Moody commutation relations. It is important to note that this
representation is only appropriate for the two-channel model as it leads to
a level-2 algebra. It would be {\it inappropriate} for the single-channel
Kondo model where the corresponding spin current generates a level-1 algebra.

In a similar way, we can also introduce Majorana representations for the
flavor currents 
\begin{eqnarray}
&&J_f^x(x)=-i\chi _2^{\prime }(x)\chi _3^{\prime }(x),  \nonumber \\
&&J_f^y(x)=-i\chi _3^{\prime }(x)\chi _1^{\prime }(x),  \nonumber \\
&&J_f^z(x)=-i\chi _1^{\prime }(x)\chi _2^{\prime }(x),
\end{eqnarray}
which reproduces the commutation relations satisfied by the flavor currents,
and $J_c(x)=-2i\chi _4^{\prime }(x)\chi _5^{\prime }(x)$ representing the
charge current operator. Note that $\chi _\alpha ^{\prime }$ with $\alpha
=1,2,3,4,5$ are also left-moving free Majorana fermion fields. It is
well-known that the dynamics of charge, flavor, and spin are completely
determined by the commutation relations of these current operators. Though
the spin currents of the two-channel Kondo model can be represented in terms
of three Majorana fermion fields $\chi _\alpha (x)$ ($\alpha =1,2,3$), they
can not be given any simple physical interpretation in terms of the original
conduction electrons $\psi _{j,\sigma }(x)$.

Now using these current operators, the Hamiltonian (2) is presented as a
quartic form in the Majorana fields. This form is convenient if one pursues
a purely algebraic approach as in the CFT \cite{al}. However, for our
purpose it is more convenient to perform an {\it inverse} Sugawara
construction using again the point-splitting procedure, and rewrite the
terms quartic in the Majorana fermions as kinetic energy terms which are
quadratic \cite{kz,ginsparg}: 
\begin{eqnarray}
&:&J_c(x)J_c(x):=4\sum_{\alpha =4}^5\chi _\alpha ^{\prime }(i\partial
_x)\chi _\alpha ^{\prime }(x);  \nonumber \\
&:&\vec{J}_f(x)\cdot \vec{J}_f(x):=2\sum_{\alpha =1}^3\chi _\alpha ^{\prime
}(i\partial _x)\chi _\alpha ^{\prime }(x);  \nonumber \\
&:&\vec{J}_s(x)\cdot \vec{J}_s(x):=2\sum_{\alpha =1}^3\chi _\alpha
(i\partial _x)\chi _\alpha (x).
\end{eqnarray}
The model Hamiltonian is thus transformed into the following two parts \cite
{zhb}, 
\begin{eqnarray}
&&H_c+H_f=\frac{v_f}{4\pi }\sum_{\alpha =1}^5\int_{-\infty }^{+\infty }dx%
\text{ }\chi _\alpha ^{\prime }(x)(i\partial _x)\chi _\alpha ^{\prime }(x), 
\nonumber \\
&&H_s=\frac{v_f}{4\pi }\sum_{\alpha =1}^3\int_{-\infty }^{+\infty }dx\text{ }%
\chi _\alpha (x)(i\partial _x)\chi _\alpha (x)  \nonumber \\
&&\text{ \quad }-\frac{iJ}2\int_{-\infty }^{+\infty }dx\text{ }\delta (x)%
\text{ }\vec{S}_d\cdot \vec{\chi}(x)\times \vec{\chi}(x).
\end{eqnarray}
$H_c+H_f$ describes the non-interacting charge and flavor degrees of
freedom. It is invariant under the symmetry group $U(1)\otimes SU(2)_2$
described by the five free Majorana fermion fields $\chi _\alpha ^{\prime
}(x)$ ($\alpha =1,2,3,4,5$). $H_s$ is the main part of the model and it
describes the spin degrees of freedom with three left-moving Majorana
fermion fields $\chi _\alpha $ ($\alpha =1,2,3$) interacting with the
impurity spin. It has the $SU(2)_2$ symmetry so that the full Hamiltonian is
described by eight different Majorana fermion fields.

In the two-channel model Hamiltonian, $H_s$ will give rise to the essential
low-energy physics of the model because it is the only part which includes
the interaction. The reduced partition function of the model Hamiltonian
with $H_s$ can be written in the form of a functional integral: 
\begin{eqnarray}
Z &=&\oint D\widehat{\Omega }\int \prod_{\alpha =1}^3D\chi _\alpha \exp
\left\{ iS_d\omega \left( \widehat{\Omega }\right) -\int_0^\beta d\tau
\right.  \nonumber \\
&&\left. \left( \int_{-\infty }^{+\infty }dx\sum_{\alpha =1}^3\frac 12\chi
_\alpha \partial _\tau \chi _\alpha +H_s\left[ \widehat{\Omega },\chi
_\alpha \right] \right) \right\} ,
\end{eqnarray}
where the impurity spin part has been expressed in terms of a spin coherent
state path integral \cite{auerbach}, $\widehat{\Omega }=(\theta ,\phi )$ is
a unit vector describing the family of spin states $|S_d,m_d\rangle $, the
eigenstates of $S_d^2$ and $S_d^z$ with eigenvalues $S_d(S_d+1)$ and $m$,
respectively, and the periodic boundary condition is assumed for the spin
vector variable.

\begin{equation}
iS_d\omega \left( \widehat{\Omega }\right) =iS_d\int_0^\beta d\tau (1-\cos
\theta )\stackrel{\cdot }{\phi }
\end{equation}
is known as the Berry phase of the spin history, which is a purely geometric
factor and will play no essential role here. In the path-integral
representation, the reduced Hamiltonian $H_s$ is expressed as 
\begin{eqnarray}
&&H_s\left[ \widehat{\Omega },\chi _\alpha \right] =\frac{v_f}{4\pi }%
\sum_{\alpha =1}^3\int_{-\infty }^{+\infty }dx\text{ }\chi _\alpha (x,\tau
)(i\partial _x)\chi _\alpha (x,\tau )  \nonumber \\
&&\hspace{1cm}-\frac{iJS_d}2\int_{-\infty }^{+\infty }dx\text{ }\delta (x)%
\text{ }\widehat{\Omega }(\tau )\cdot \vec{\chi}(x,\tau )\times \vec{\chi}%
(x,\tau ),
\end{eqnarray}
where $\chi _\alpha (x,\tau )$ ($\alpha =1,2,3$) are real Grassmann
variables corresponding to the three Majorana fermion fields and as far as $%
\chi _\alpha $ are concerned, the path integrals over them are bilinear. The
partition function can thus be rewritten in the following form 
\begin{eqnarray}
Z &=&\oint D\widehat{\Omega }\exp \left( iS_d\omega \left( \widehat{\Omega }%
\right) \right) \int \prod_{\alpha =1}^3D\chi _\alpha   \nonumber \\
&&\exp \left\{ -\frac 12\int_0^\beta d\tau \int_{-\infty }^{+\infty }dx\text{
}\Psi ^{\dagger }(x,\tau )\widehat{M}\text{ }\Psi (x,\tau )\right\} ,
\end{eqnarray}
with a three-component vector $\Psi ^{\dagger }(x,\tau )=(\chi _1(x,\tau ),$ 
$\chi _2(x,\tau ),$ $\chi _3(x,\tau ))$ and its transposition $\Psi (x,\tau
).$ The matrix is denoted by $\widehat{M}=(\partial _\tau -\overline{v}%
_fi\partial _x)I+iJS\delta (x)\widehat{M^{\prime }}(\tau ),$ where 
\begin{equation}
\widehat{M^{\prime }}(\tau )=\left[ 
\begin{array}{ccc}
0, & -\Omega ^z(\tau ), & \Omega ^y(\tau ) \\ 
\Omega ^z(\tau ), & 0, & -\Omega ^x(\tau ) \\ 
-\Omega ^y(\tau ), & \Omega ^x(\tau ), & 0
\end{array}
\right] ,
\end{equation}
$\overline{v}_f=v_f/2\pi ,$ and $I$ is a $3\times 3$ unit matrix. Then we
integrate out the variables $\chi _\alpha $ and obtain an effective action
which only contains the spin vector variables: 
\begin{eqnarray}
Z &=&Z_0\oint D\widehat{\Omega }\exp \left( iS_d\omega \left( \widehat{%
\Omega }\right) -S_{eff}\right) ,  \nonumber \\
S_{eff} &=&-\frac 12\text{Tr }\ln \left[ 1+iJS_d\delta (x)\widehat{G}%
\widehat{M^{\prime }}(\tau )\right] ,
\end{eqnarray}
where $Z_0=\frac 12\det \left[ (\partial _\tau -\overline{v}_fi\partial
_x)I\right] $ is the partition function of the non-interacting limit of $H_s,
$ and its free Majorana fermion propagator is given by $\widehat{G}%
=(\partial _\tau -\overline{v}_fi\partial _x)^{-1}$. Tracing here is taken
over space, imaginary time, and the matrix indices. Using the identity 
\[
\text{Tr }\ln (1+\widehat{A})=-\sum_{n=1}^\infty \frac{(-1)^n}n\text{Tr }(%
\widehat{A})^n,
\]
we obtain the general expression for the effective action 
\begin{eqnarray}
S_{eff} &=&\sum_{n=1}^\infty \frac{(-iJS_d)^n}{2n}\int_0^\beta d\tau
_1...\int_0^\beta d\tau _nG(\tau _{12})G(\tau _{23})...  \nonumber \\
&&\text{ \quad }G(\tau _{n1})\text{Tr}\left[ \widehat{M^{\prime }}(\tau _1)%
\widehat{M^{\prime }}(\tau _2)......\widehat{M^{\prime }}(\tau _n)\right] ,
\end{eqnarray}
where the space integration has been easily carried out due to the presence
of the delta function so that the free Majorana fermion propagators are
replaced by the local one, $G(\tau )=$ $\frac 1{\overline{v}_f}\frac{\pi
/\beta }{\sin (\pi \tau /\beta )}$ , and $\tau _{ij}=\tau _i-\tau _j$. Note
that $\tau _1,\tau _2,....,\tau _n$\ are {\it unequal} imaginary times. In
fact $S_{eff}^{(n)}$ is the contribution of a one-loop diagram made of $n$
local propagators $G(\tau _{ij})$ and $n$ local vertices $iJS\widehat{%
M^{\prime }}(\tau _i)$. The factor $1/(2n)$ in front of each term is due to
symmetry of the corresponding diagram. In the above derivation, the impurity
spin is not specified to be $1/2$ and only the spin of the conduction
electrons has been assumed to $1/2$. However, the following discussion will
focus on the overscreening case $\left( S_d=1/2\right) $.

Up to second order of $JS_d,$ the effective action is worked out as 
\begin{equation}
S_{eff}^{(2)}=\frac 12(JS_d)^2\int_0^\beta d\tau _i\int_0^\beta d\tau
_jG^2(\tau _{ij})\widehat{\Omega }(\tau _i)\cdot \widehat{\Omega }(\tau _j),
\end{equation}
which describes a Heisenberg spin chain with an inverse-square long-range 
{\it antiferromagnetic} interaction between the impurity spins at two
different imaginary times, and the sign of the original Kondo exchange
coupling (ferromagnetic or antiferromagnetic) is not distinguishable here.
Note that the spin coherent state path integral has assumed a periodic
boundary condition for the spin variable so that the Heisenberg spins in
fact sit on a circle of length $\beta $ with exchange inversely proportional
to the square of the distance between spins, which has the same form of the
path-integral functional as the one-dimensional Heisenberg spin chain with
an inverse-square long-range interaction, the HS model \cite{hs} in the {\it %
semiclassical} limit. Therefore, all the {\it static} properties of the HS
model can be readily translated to the present model. (a) The low-energy
states of $S_{eff}^{(2)}$ in the large-$\beta $ limit are described by the
chiral-SU(2) invariant $k=1$ Wess-Zumino-Witten model, which is a
conformally invariant Gaussian field theory. $S_{eff}^{(2)}$ thus can
represent a fixed point action of the two-channel Kondo model, and the
elementary excitations are spinons, i.e., the $S=1/2$ particles instead of
spin waves which are the elementary excitations of an ordered
antiferromagnet. The spinons satisfy semion statistics intermediate between
bosons and fermions, being an example of the {\it exclusion} statistics
interpretation of fractional statistics \cite{haldane}.(b) $S_{eff}^{(2)}$
describes a free gas of spinons, being a fundamental model for gapless spin-$%
1/2$ antiferromagnetic spin model, and the dominant asymptotic spin-spin
correlation function of $S_{eff}^{(2)}$ is algebraic with an universal
exponent $\eta =1$ without{\it \ }logarithmic corrections \cite{hs}.

\begin{equation}
<T_\tau \widehat{\Omega }(\tau _i)\cdot \widehat{\Omega }(\tau _j)>\sim 
\frac 1{\mid \tau _i-\tau _j\mid }
\end{equation}
and the impurity spin variable $\widehat{\Omega }(\tau )$ thus acquires a 
{\it dynamic} scaling dimension $1/2.$ Such a behavior is also the universal
spin-spin correlation function of the low-energy non-FL behavior of the
two-channel Kondo model \cite{al,ek,ml,cit}, leading to a marginal FL form 
\cite{varma} of the impurity spin spectrum: Im$\chi _d(\omega +i0^{+})\sim
\tanh (\frac \omega {2T}).$ (c). We can thus conclude that $S_{eff}^{(2)}$
represents the low-energy non-FL fixed point action of the spin part of the
two-channel Kondo model.

The third order of $JS_d$ can be viewed as a correction to the fixed point
action, which has been derived as

\begin{eqnarray}
S_{eff}^{(3)} &=&\frac{-i}6(JS_d)^3\int_0^\beta d\tau _i\int_0^\beta d\tau
_j\int_0^\beta d\tau _k\text{ }  \nonumber \\
&&\text{ }G(\tau _{ij})G(\tau _{jk})G(\tau _{ki})\widehat{\Omega }(\tau
_i)\cdot \left( \widehat{\Omega }(\tau _j)\times \widehat{\Omega }(\tau
_k)\right) .
\end{eqnarray}
This term describes a Heisenberg spin chain with a long-range interaction of
the impurity spins at three different imaginary times and is completely
antisymmetric in its indices ($ijk$). In accordance with the low-energy
non-FL fixed point action $S_{eff}^{(2)}$, this interaction term is
irrelevant because it has a dynamic scaling dimension $3/2.$ However, it is
this interaction that distinguishes the sign of the original Kondo exchange
coupling, so that it is a {\it dangerous} irrelevant operator. It is
conceivable that the non-FL thermodynamic properties of the two-channel
Kondo model around the low-energy fixed point are derived from a
perturbation theory of $S_{eff}^{(3)}$. For instance, the second order
perturbation calculation of $S_{eff}^{(3)}$ gives rise to the extra
low-temperature specific heat due to the presence of the impurity spin as $%
T\ln T$ . The detailed calculations of the thermodynamic properties will be
given in the future publication.

When the expansion is carried out to the fourth order, we obtain a four-spin
long-range interaction of the impurity spins at four different times 
\begin{eqnarray}
&&S_{eff}^{(4)}=\frac 14(JS_d)^4\int_0^\beta d\tau _i\int_0^\beta d\tau
_j\int_0^\beta d\tau _k\int_0^\beta d\tau _l \\
&&\text{ \quad }G(\tau _{ij})G(\tau _{jk})G(\tau _{kl})G(\tau _{li})\widehat{%
\Omega }(\tau _i)\cdot \widehat{\Omega }(\tau _j)\text{ }\widehat{\Omega }%
(\tau _k)\cdot \widehat{\Omega }(\tau _l),  \nonumber
\end{eqnarray}
which is clearly irrelevant as far as the low-energy fixed point is
concerned, as its dynamic scaling dimension is $2.$ All higher-order terms
thus contain no essential physics and can be neglected completely.
Therefore, the reduced effective action will be given by $%
S_{eff}=S_{eff}^{(2)}+S_{eff}^{(3)}$, which is a natural generalization of
the inverse-square long-range HS spin exchange model.

In summary, we use the Majorana fermion representation of the algebra of
spin, charge, and flavor currents to transform the two-channel Kondo model.
In a path integral formulation, we derive a reduced effective action of the
two-channel Kondo model, which is a one-dimensional Heisenberg spin chains
with two-spin, three-spin, etc. long-range interactions, as a natural
generalization of the inverse-square long-range HS spin model. It is argued
that the nontrivial two-channel Kondo physics in the low-energy regime can
be reproduced from the first two terms of the spin action, and the other
relevant issues are under investigation. As pointed out in Ref. \cite{fdm},
the infinite set of multiplicative degeneracy of HS model\cite{haldane} is
due to the hidden $SU(2)$ Yangian symmetry. Comparing our present
formulation with the CFT treatment of the two-channel Kondo problem\cite{al}%
, this statement becomes apparent. After completing the present work, we
become aware of a general review article \cite{schlottmann2} on exact
results for highly correlated electron systems in one dimension, where some
analogies of the inverse square long-range models to other interacting
models are discussed within the framework of the Bethe-Ansatz.

{\bf Acknowledgment}

One of the authors (G. -M. Zhang) would like to thank A. C. Hewson for his
helpful discussions on the two-channel Kondo model.

\end{document}